\documentclass[a4paper,10pt,twoside]{cpc-hepnp}

\usepackage{multicol}
\usepackage{graphicx}
\usepackage{booktabs}
\usepackage{amssymb,bm,mathrsfs,bbm,amscd}
\usepackage[tbtags]{amsmath}
\usepackage{lastpage}

\usepackage{multirow}

\begin{document}

\fancyhead[c]{\small Submitted to ¡®Chinese Physics C'}
\fancyfoot[C]{\small 010201-\thepage}

\title{The study of neutron activation yields in spallation reaction of 400 MeV/u carbon on a thick lead target\thanks{Supported by National Natural Science
Foundation of China (11105186, 11105187) and the West Light Foundation of the Chinese Academy of Sciences (XBB100123) }}

\author{%
      MA Fei$^{1}$
\quad GE Hong-Lin$^{1;1)}$\email{mf@impcas.ac.cn}%
\quad ZHANG Xue-Ying$^{1}$
\quad ZHANG Hong-Bin$^{1}$\\
\quad JU Yong-Qin$^{1}$ \quad CHEN Liang$^{1}$ \quad YANG
Lei$^{1}$
\quad FU Fen$^{1}$\\
\quad ZHANG Ya-Ling$^{1}$ \quad LI Jian-Yang$^{1}$ \quad LIANG
Tian-Jiao$^{2}$
\quad ZHOU Bin$^{3}$\\
\quad WANG Song-Lin$^{3}$ \quad LI Jin-Yang$^{1}$ \quad XU
Jun-Kui$^{1}$
\quad LEI Xiang-Guo$^{1}$\\
\quad QIN Zhi$^{1}$ \quad GU Long$^{1}$ \quad JIN Gen-Ming$^{1}$ }
\maketitle

\address{%
$^1$ Institute of Modern Physics, Chinese Academy of Sciences, Lanzhou 730000, China\\
$^2$ Institute of Physics, Chinese Academy of Sciences, Beijing 100190, China\\
$^3$ Institute of High Energy Physics, Chinese Academy of
Sciences, Beijing 100049, China }

\begin{abstract}
The spallation-neutron yield was studied experimentally by
bombarding a thick lead target with 400 MeV/u carbon beam. The
data were obtained with the activation analysis method using foils
of Au, Mn, Al, Fe and In. The yields of produced isotopes were
deduced by analyzing the measured ${\gamma}$ spectra of irradiated
foils. According to the isotopes yields, the spatial and energy
distributions of the neutron field were discussed. The
experimental results were compared with Monte Carlo simulations
performed by the GEANT4 + FLUKA code.
\end{abstract}



\begin{multicols}{2}

\section{Introduction}

Spallation reactions can be used to produce intense neutron fluxes
with a high energy beam on a thick target with high atomic number.
Recently, the possible applications are rapidly growing in many
fields, such as spallation neutron source (SNS) and accelerator
driven system (ADS)~\cite{lab1,lab2}. For designing the spallation
target and shielding of accelerator facilities, it is necessary to
estimate the production and distribution of the spallation
reaction products especially for the neutrons. Although the
related researches have been carried out in Europe, USA, Japan and
China for many years~\cite{lab3}, there are still needs to
accumulate more experiment data to test the applicability of all
kinds of model descriptions. As a part of a complex research of
SNS and ADS in China, we have studied the spallation reaction by
irradiating a lead target with a high energy carbon beam. In this
work, the spallation-neutron field was measured based on the
activation analysis method~\cite{lab4}, and the experimental
results were compared with GEANT4 plus FLUKA
simulations~\cite{lab5,lab6}.

\section{Experiment setup}

The experiment was performed at the HIRFL-CSR in Lanzhou, China. A
$^{12}$C$^{6+}$ beam with energy of 400 MeV/nucleon was used to
bombard a massive cylindrical lead target. The diameter and total
length of the target were 10 cm and 25 cm, respectively. The
average beam intensity was 1.72¡Á10$^{7}$ pps as monitored by the
proportional chamber. The irradiation of the beam continued for 24
hours, and the course of the irradiation is shown in Fig 1. The
activation foils of Al, Au, Mn, Fe and In were placed on the
surface of the target so as to measure the leakage neutrons. The
foils of each group were located at three longitudinal distances
of 1.0, 5.0 and 11.0 cm from upstream surface of the target, see
Fig 2. The detailed description for the foils is summarized in
Table. 1. Two groups of Au foils were arranged on two opposite
sides of the target and used to determine the position deviation
of the beam. Via reactions (n, ${\gamma}$), (n, xn) and (n, xn,
yp), the stable isotopes composing of the detector foils were
transmuted into radioactive ones, which were identified by
observing the characteristic ${\gamma}$ rays. In order to
determine the isotopes with different half-lives, each irradiated
foil was measured for several times by one HPGe detector with
relative efficiency of 65 \% and energy resolution of 1.90 keV at
1.33 MeV. The distance between the detector endcap and the foils
was 3.0 cm. The detector efficiency was calibrated with the
standard point-like sources $^{60}$Co, $^{133}$Ba, $^{137}$Cs and
$^{152}$Eu.

\begin{center}
\includegraphics[width=8cm]{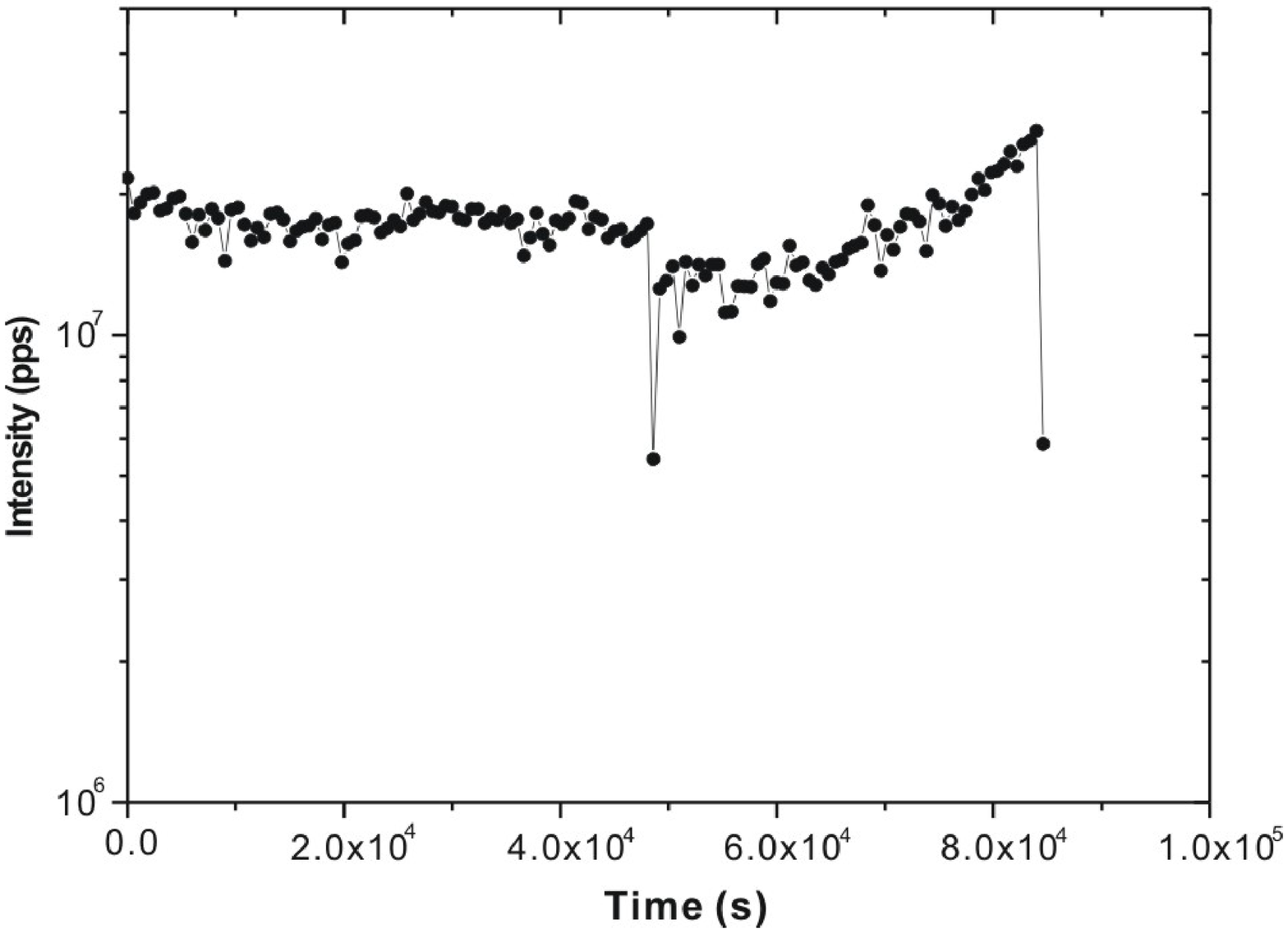}
\figcaption{\label{Fig1}   Course of irradiation with 400 MeV/u
carbons. }
\end{center}

\begin{center}
\includegraphics[width=8cm]{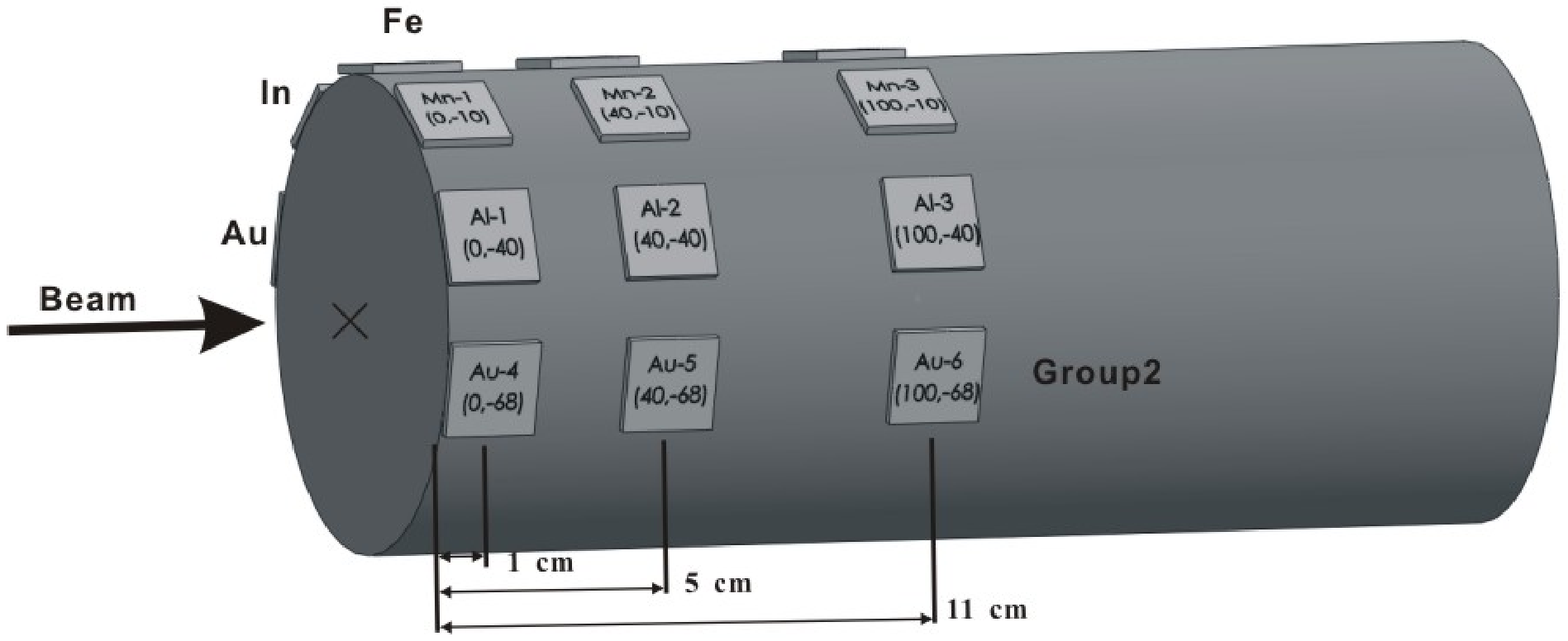}
\figcaption{\label{Fig2}   Scheme of placement of activation
foils. }
\end{center}

\end{multicols}

\begin{center}
\tabcaption{ \label{tab2}  The size of activations foils and the
observed reactions, their thresholds, and half-lives of products.}
\footnotesize
\begin{tabular*}{170mm}{@{\extracolsep{\fill}}ccccccc}
\toprule
 Activation Foil & Abundance(\%) & Area (mm$^{2}$) & Thikcness(mm) & Reaction & Threshold   energy (MeV) & Half-life \\
\hline
\multirow{3}{*}{$^{197}$Au} & \multirow{3}{*}{99.99} & \multirow{3}{*}{20¡Á20} & \multirow{3}{*}{0.1} & $^{197}$Au(n, ${\gamma}$)$^{198}$Au & - & 2.69517 d \\
\cline{5-7}
& & & & $^{197}$Au(n, 2n)$^{196}$Au & 8.1 & 6.1669 d\\
\cline{5-7}
& & & & $^{197}$Au(n, 4n)$^{194}$Au & 24 & 1.584 d\\
\hline
$^{27}$Al&99.99&20¡Á20&2.0&$^{27}$Al(n, ${\alpha}$)$^{24}$Na&4.6&14.959 h\\
\hline
\multirow{3}{*}{$^{55}$Mn} & \multirow{3}{*}{99.99} & \multirow{3}{*}{20¡Á20} & \multirow{3}{*}{3.0} & $^{55}$Mn(n, ${\gamma}$)$^{56}$Mn & - & 2.5785 h \\
\cline{5-7}
& & & & $^{55}$Mn(n, 2n)$^{54}$Mn & 10.5 & 312.12 d\\
\cline{5-7}
& & & & $^{55}$Mn(n, 4n)$^{52}$Mn & 31.4 & 5.591 d\\
\hline
\multirow{2}{*}{$^{56}$Fe} & \multirow{2}{*}{91.71} & \multirow{2}{*}{20¡Á20} & \multirow{2}{*}{2.0} & $^{56}$Fe(n, p)$^{56}$Mn &3 & 2.5785 h \\
\cline{5-7}
& & & & $^{56}$Fe(n, t)$^{54}$Mn & 12.2 & 312.12 d\\
\hline
\multirow{4}{*}{$^{115}$In} & \multirow{4}{*}{95.69} & \multirow{4}{*}{20¡Á20} & \multirow{4}{*}{2.0} & $^{115}$In(n, ${\gamma}$)$^{116m}$In & - & 54.29 m \\
\cline{5-7}
& & & & $^{115}$In(n, n')$^{115m}$In & 0.35 & 4.486 h\\
\cline{5-7}
& & & & $^{115}$In(n, 2n)$^{114m}$In & 9.1 & 49.51 d\\
\cline{5-7}
& & & & $^{115}$In(n, 5n)$^{111}$In & 33.5 & 2.807 d\\

\bottomrule
\end{tabular*}%
\end{center}

\begin{multicols}{2}

\section{Result and Discussions}

The measured ${\gamma}$ spectra were processed by the GAMMA-W code
which was applied to calculate the net ${\gamma}$-peak areas via
an unfolding algorithm using a least-squares fit ~\cite{lab7}.
Considering the decay during irradiation, cooling and measurement,
the activation yields R (i.e., number of activated nuclei per
nucleus of the activated foil and per one incident carbon) of the
corresponding radioactive nuclei could be determined according
 to the relation
\begin{equation}
\label{eq1}
R = \frac{c \lambda e^{\lambda} t_{d}}{\epsilon_{\gamma} I_{\gamma} N_{t} f_{s} I_{c} (1-e^{-\lambda t_{c}}) (1-e^{-\lambda t_{irr}})}
\end{equation}

where C is fitted area of the ${\gamma}$-peak; I$_{\gamma}$ is
intensity of this ${\gamma}$ transition per decay; f$_{s}$ is
self-absorption correction factor of ${\gamma}$ transition;
I$_{c}$ is average beam intensity; ${\epsilon}$$_{\gamma}$ is
detector efficiency; N$_{t}$ is nueclear numbers of the foil;
${\lambda}$ = ln2/T$_{1/2}$ is disintegration
 constant; t$_{d}$ is time from the end of irradiation to the beginning of the measurement; t$_{irr}$ is
  time of irradiation; t$_{c}$ is time of measurement. As a function of the position along the target,
  the activation yields produced in Al, Au, Mn, Fe and In foils are shown in Fig 3. Errors in
  the figure concern the statistical error, the coincidence summing effect of ${\gamma}$ transition,
  inaccuracy of the I$_{\gamma}$ and ${\epsilon}$$_{\gamma}$. Systematic error, such as the inaccuracy of the beam intensity
  and direction, contributes about 10 \%.

Because of the large size of the target ($\Phi{100}\times{250}$)
as well as the small beam section (FWHM=2.5 cm), it can be
believed that most of the radioactive nuclei in the foils are
induced by the emitted neutrons but the beam or other spallation
products. As shown in Fig 3, it is found that the productions of
$^{198}$Au, $^{116}$In and $^{56}$Mn (in Mn foils), which are
produced in (n, ${\gamma}$) reactions, have no major difference
along the longitudinal direction. The reason could be the quite
homogeneous field of thermal neutrons on the surface of the target
that give constant contribution to the productions in different
positions. The homogeneous field is mainly produced in two
processes. One is the evaporation phase in spallation reaction
where the produced thermal neutrons have nearly isotopes angular
distributions~\cite{lab3}. On the other hand, the leakage neutrons
are multi-scatted by the lab equipments surrounding the spallation
target which also lead to the homogeneous field of thermal
neutrons as a background. For the isotopes produced in threshold
reactions, the activation yields at the 1 cm and 11 cm are more or
less the same and both lower than the yields at 5 cm. Therefore,
we can conclude that the intensity of the fast neutron field at
the position of 5 cm is higher than that at other two positions.

\begin{center}
\includegraphics[width=8cm]{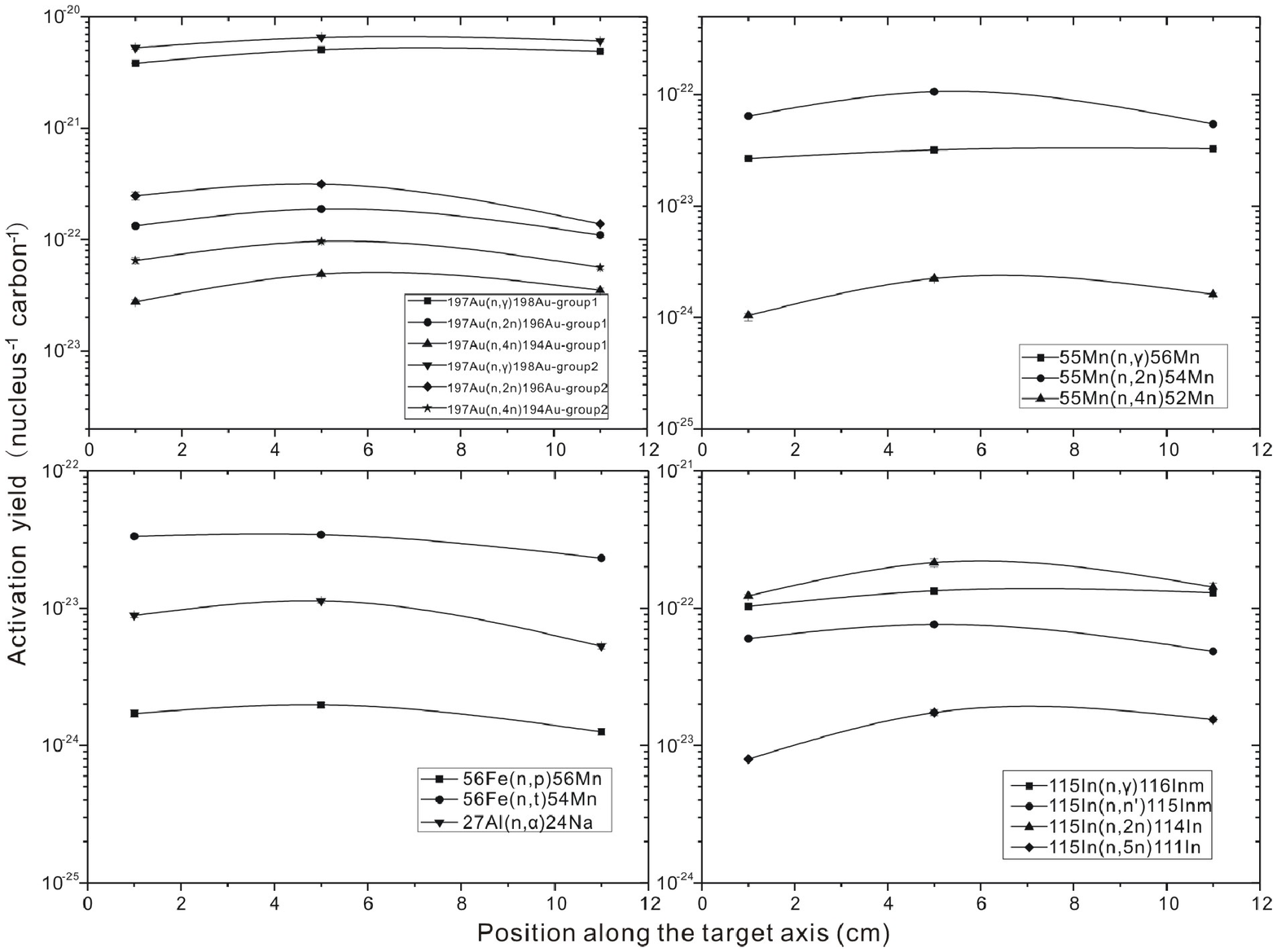}
\figcaption{\label{Fig3}   Longitudinal distributions of
activation yields of the foils. }
\end{center}

\begin{center}
\includegraphics[width=8cm]{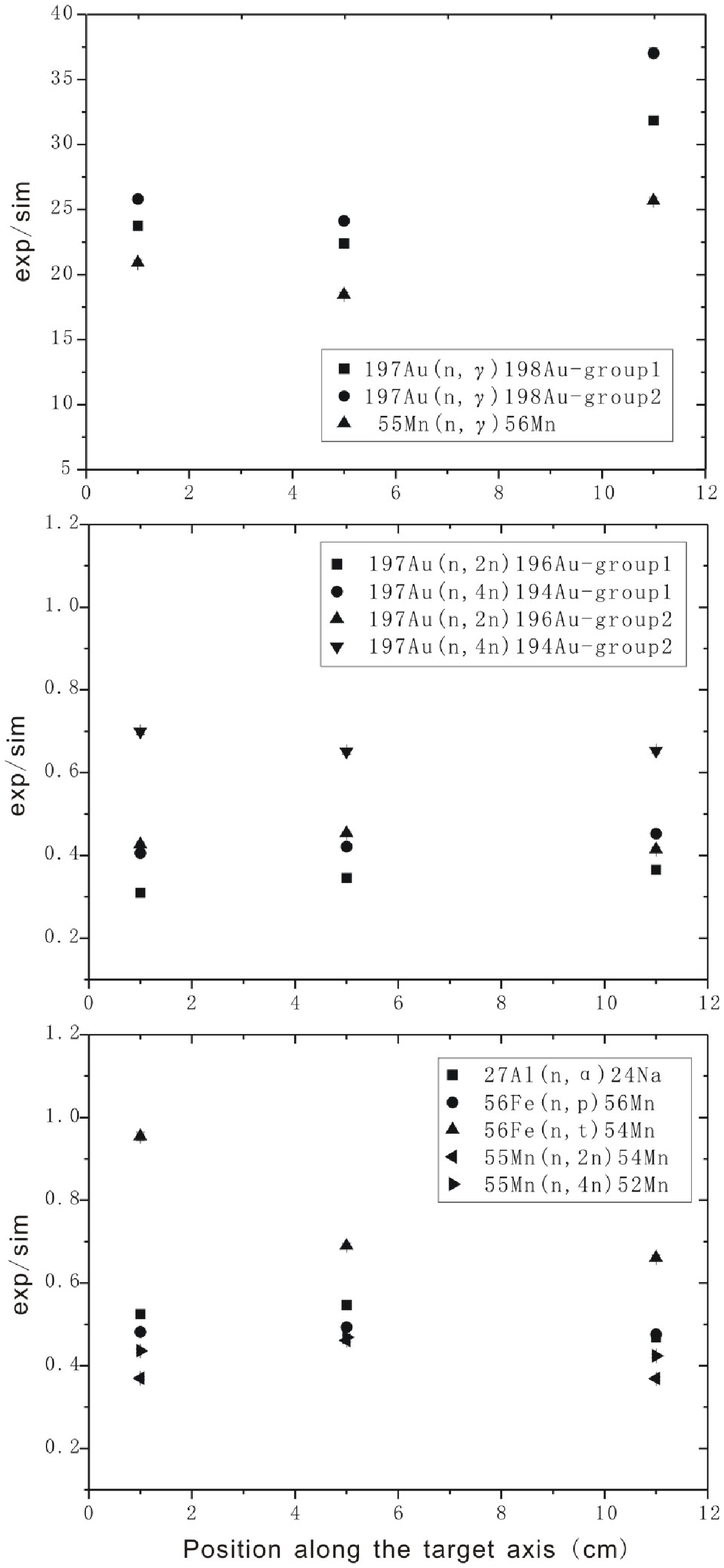}
\figcaption{\label{Fig4}   GEANT4 simulations of neutron spectra
at different longitudinal positions. }
\end{center}

\begin{center}
\includegraphics[width=6cm]{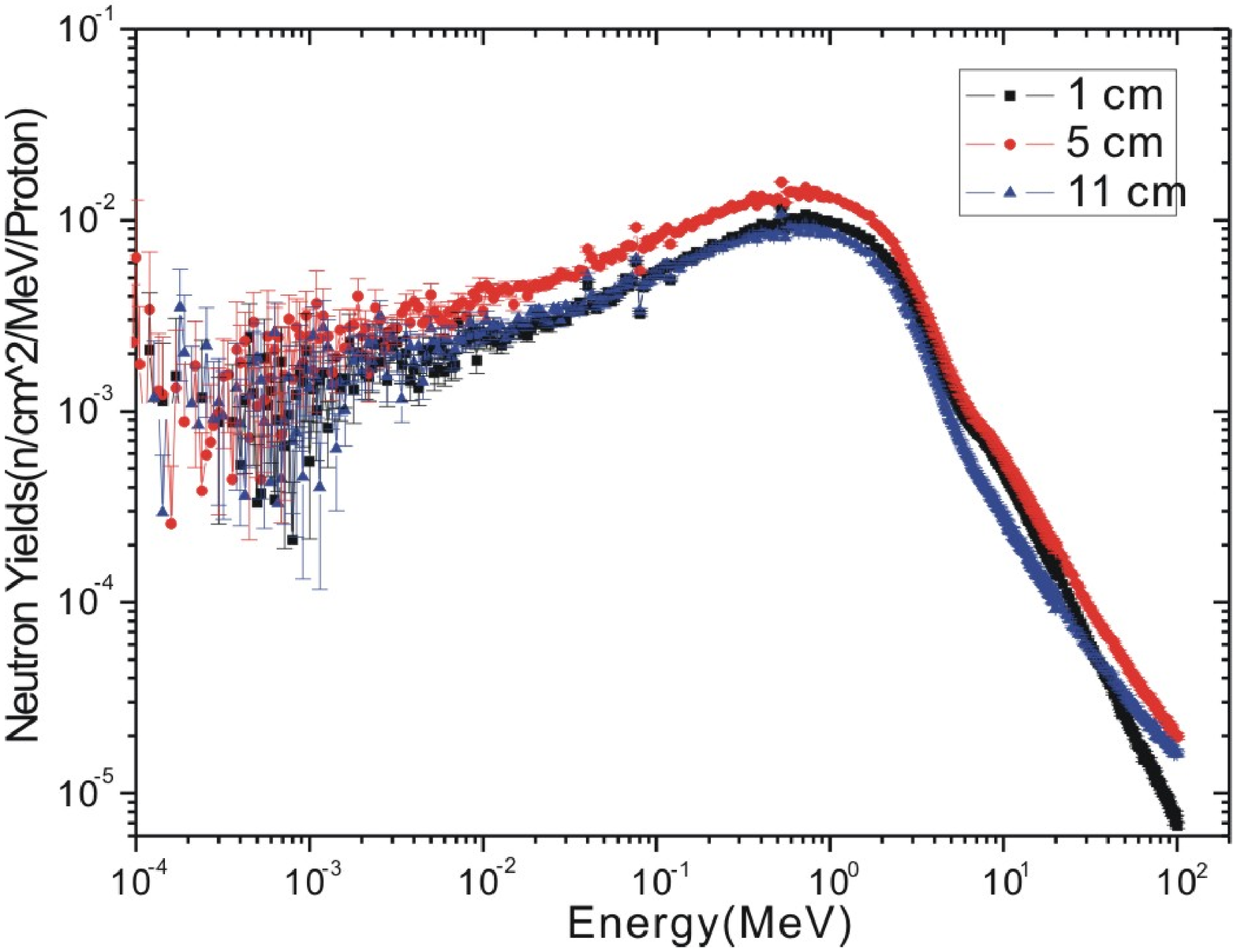}
\figcaption{\label{Fig5}   Comparison of the experimental yields
of activated isotopes versus the yields from the GEANT4 plus FLUCA
simulation.}
\end{center}

\section{Comparison Between Simulation and Experiment}

The processing of the experimental data was accompanied by
simulations of the neutron production and the activation yields.
The simulation of the spallation reaction was performed by GEANT4.
The Intranuclear cascade stage and the Equilibrium stage were
consisted in the simulation of the course. The simulated energy
spectra are shown in Fig 4. As seen in the figure, the maximum
intensity of the neutron field emitted from target is located at
the 5 cm. The energetic spectrum at the end of the target is
similar to the simulation at the beginning. Qualitatively, we drew
the same conclusions from the experimental results, see Fig 3.
Combining the simulated neutron spectra, the activation yields of
the foils were calculated by FLUKA. Those default FLUKA neutron
cross sections for energy below 20 MeV were generally taken from
ENDF/B-VIIRO library~\cite{lab8}, for higher energy were mostly
calculated by FLUKA itself. The activation yields of In foils were
not calculated because of that FLUKA couldn¡¯t give the proportion
of the isomer in the isotope production. The comparison of
activation yields between experimental data with the calculated
value is shown in Fig 5. For the neutron capture reactions, the
large ratios indicate that the experimental values are much
greater than simulated results. The reason should be attributed to
that the simulations substantially underestimate the contribution
of background to the thermal neutron field. In the contrary, the
ratios for threshold reactions are in the region of 0.3 to 1.0. In
other words, the local discrepancies are as small as a factor of 3
in extreme cases. This indicates that GEANT4 plus FLUKA codes
could produce overall yield magnitudes quite well for fast
neutrons.

\section{Conclusion}

In summary, we have studied the neutron production in the reaction
of high energy carbons bombarding a thick lead target. The
intensity and distribution of neutron field were measured by the
neutron activation analysis method. By analyzing the activation
yields in different positions, it is found that the maximum
intensity of the fast neutron field produced in the spallation
target was located in the position of 5 cm from the target
forehead. Different from the former, the homogeneous field of
thermal neutron was measured on the surface of the target. We also
compared the experiment data with the simulations. It was shown
that the calculations were in agreement with the experimental data
in magnitudes for fast neutron production, and it should be
emphasized that the simulations for neutron capture reactions must
value the contribution of lab setup.

\end{multicols}

\vspace{10mm}

\vspace{-1mm}
\centerline{\rule{80mm}{0.1pt}}
\vspace{2mm}

\begin{multicols}{2}

\end{multicols}

\clearpage

\end{document}